\documentclass[12pt]{article} 
\usepackage[koi8-r]{inputenc}
\usepackage[english]{babel}
\usepackage{pazh}
\usepackage{graphicx,amssymb,amsmath}
\usepackage{natbib}
\usepackage{xspace}
\tightenlines

\newcommand{\Teff}{$T_{\rm eff}$}
\newcommand{\lgg}{log\,$g$}

\newcommand{\eps}[1]{\log\varepsilon_{\rm #1}}
\newcommand{\kms}{${\rm km s^{-1}}$}
\newcommand{\kH}{$S_{\!\rm H}$}    %%% Note negative spaces!
\newcommand{\Eexc}{$E_{\rm exc}$}

\newcommand{\eu}[5]{\mbox{$#1\,^#2{\rm #3}^{#4}_{\rm #5}$}}

\sloppypar

\begin{document}
	
\baselineskip 21pt

\title{\bf Non-LTE corrections for determinations of europium abundances in F-G-K stars in a broad metallicity range}

\author{\bf \hspace{-1.3cm}\copyright\, 2024 \ \
L.~I.~Mashonkina\affilmark{1*}, S.~A.~Yakovleva\affilmark{2} }

\affil{
{\it Institute of Astronomy, RAS, Pyatnitskaya st. 48, 119017 Moscow, Russia$^1$ \\
 Herzen State Pedagogical University, Kazanskaya ul. 5, 191186 St. Petersburg, Russia$^2$}
}

	\vspace{2mm}
	
	\sloppypar 
	\vspace{2mm}
	
Europium plays a key role in studies of nucleosynthesis in the rapid (r-) process of neutron capture nuclear reactions and the evolution of the r-process element abundances in galaxies. We refine the method for analyzing the Eu~II lines in stellar spectra by updating the Eu~II model atom with recent data on the rate coefficients for inelastic processes in the Eu~II + H~I collisions. The method was tested by deriving abundances from the Eu~II 4129~\AA\ and 6645~\AA\ lines in the Sun and five reference stars with well determined atmospheric parameters and high-resolution spectra available. It was shown that abandoning the local thermodynamic equilibrium (LTE) assumption allows the abundances from the two lines to be matched within the error bars, while the difference in the LTE abundances amounts to $-0.09$~dex for the Sun and $-0.07$~dex to $-0.18$~dex for the metal-poor stars. Accounting for the departures from LTE (non-LTE effects) for Eu~II leads to reducing the statistical errors of the derived stellar Eu abundances. The non-LTE abundance corrections for 11 lines of Eu~II were calculated for a grid of the plane-parallel MARCS model atmospheres with effective temperatures, surface gravities, and metallicities relevant to F-G-K type stars. They are publicly available and can be applied for improving stellar Eu abundances in studies of the galactic chemical evolution.
	

\vfill
\noindent\rule{8cm}{1pt}\\
{$^*$ e-mail $<$lima@inasan.ru$>$}

\section{Introduction}

Throughout the Galaxy lifetime, europium has been synthesized in the rapid (r-) process of neutron capture nuclear reactions. The contribution of the r-process to the solar Eu abundance is estimated at 95.1\%\ (Prantsos~et~al., 2020). Such reactions occur at very high neutron number densities of $N_{\rm n} = 10^{20}-10^{28}$~cm$^{-3}$, which is associated with explosive conditions in stars. A review of possible astrophysical sites of the r-process is given by Cowan~et~al. (2021). One of them, the ejection of matter during the merger of two neutron stars, was observationally confirmed. The source of gravitational waves GW170817 (Abbot~et~al., 2017a) was observed in different spectral ranges, and the kilonova AT2017gfo, which arose as a result of the merger of neutron stars, was detected (Abbot~et~al., 2017b). It turned out that the bolometric light curve of AT2017gfo corresponds to the emission of a large number of radioactive isotopes synthesized in the r-process. But kilonovae cannot be the only site of the r-process, since heavy elements such as Ba and Sr are observed in stars formed in the early Galaxy, when the iron abundance had not yet reached [Fe/H]\footnote{for any two elements X and Y: [X/Y] = log$(N_{\rm X}/N_{\rm Y})_{star} - \lg (N_{\rm X}/N_{\rm Y})_{Sun}$} $= -4$, i.e. before the first explosions of kilonovae (Briel~et~al., 2022). On a short time scale, enrichment of the interstellar medium with heavy elements can be provided by the r-process in magnetorotational supernovae, which explode owing to rapid rotation and strong magnetic field of their progenitors (Leblanc and Wilson, 1970; Bisnovatyi-Kogan, 1971). Correct modeling of the r-process in such complex conditions as explosions of kilonovae and magnetorotational supernovae requires three-dimensional magnetohydrodynamic calculations \citep[see, e.g.,][]{mrsn2024}. Galactic chemical evolution models for the r-process elements are tested by determining their abundances in stars of different metallicities.

Of all the r-process elements, i.e. elements for which the r-process contribution to the solar abundance exceeds 80\%\ \citep[Gd, Tb, Dy, etc.,][]{Prantzos2020}, europium is the best observed element in the F-G-K stars, due to the fact
that the Eu~II 4129 and 4205~\AA\ resonance lines lie in the visible spectral range and are only weakly blended. In the literature, there are numerous determinations of stellar Eu abundances over a wide range of metallicities, but in most cases they were made under the simplifying assumption of local thermodynamic equilibrium (LTE), see, for example,\citet{McWilliam95,Burris00,Francois07,Hill2019}. The first analysis taking into account deviations from LTE (non-LTE effects) was performed by \cite{eunlte2000} for the Sun and a sample of stars of different metallicity. They applied the Eu~II model atom, which was built by Mashonkina (2000) using all 163 energy levels represented in the NIST database\footnote{https://physics.nist.gov/asd} (Kramida~et~al. 2026) up to \eu{4f^7 8s}{7}{S}{\circ}{} with excitation energy of \Eexc\ = 8.287~eV, but that atomic model was modified by \cite{eunlte2000}\footnote{Despite the detailed description of the atomic model in Sect.~3.1, Figure~1, Table~2 of this paper, Storm et al. (2024) erroneously claim that the atomic model of \cite{eunlte2000} is not published and assume that it includes levels only up to \eu{e}{7}{S}{\circ}{} with \Eexc\ = 6.155~eV.} by combining closely spaced levels of the same parity, so that, in the non-LTE calculations, it includes 63 levels of Eu~II and the ground state of Eu~III.

The next step in improving non-LTE calculations for Eu~II lines was made by \cite{Storm2024}. Their model atom includes the same 163 levels of Eu~II, as in the work of \cite{eunlte2000}, and additionally 498 levels of Eu~I. For
Eu~II, \cite{Storm2024} use the same data on transition probabilities and the same theoretical approximations for calculating photoionization cross sections, electron-impact excitation and ionization rates. However, unlike \cite{eunlte2000}, who used the Drawin approximation (Drawin, 1969) as presented by Steenbock and Holweger (1984) to treat collisions with hydrogen atoms, \cite{Storm2024} applied the rate coefficients for the excitation and ion pair production obtained in quantum mechanical calculations. When analyzing the Eu~II 6645~\AA\ line in the solar spectrum, \cite{Storm2024} used both a classical, plane-parallel (1D) and a three-dimensional (3D) model atmosphere based on hydrodynamic calculations. The difference in the non-LTE abundance between the 3D and 1D calculations was $-0.03$~dex. The same atomic model was then applied by \cite{Guo2025} to determine the non-LTE abundances (1DNLTE) of europium in a sample of stars in the $-2.4 \le$ [Fe/H] $\le -0.5$ range.

Until recently, non-LTE calculations for Eu~II were carried out only with the model atom from \cite{eunlte2000}, with an approximate treatment of collisions with H~I atoms according to the formulas of Steenbock and Holweger (1984), which include the scaling coefficient \kH. Europium non-LTE abundances have been determined for large stellar samples \citep[e.g.,][]{Zhao2016,mash_dsph}. Calculations of inelastic processes in the Eu~II + H~I collisions \citep{Storm2024} 
 raised the question of how large the errors in the previously obtained Eu abundances are due to the use of the Drawin approximation. The present study aimed to:
\begin{itemize}
\item Update of the Eu~II model atom by implementing the rate coefficients for excitations and ion pair formations in the Eu~II + H~I collisions obtained from quantum mechanical calculations, both our own and those available in the literature.
\item Testing the atomic model by determining the abundances from the Eu~II lines in the Sun and selected stars with known atmospheric parameters.
\item Calculations of the non-LTE abundance corrections for the Eu~II lines in a wide range of stellar parameters, which will be publicly available and can be used in future studies of the Eu abundance evolution in galaxies.
\end{itemize}

The article is structured as follows. Section~2 describes the methodology for calculating the rate coefficients for inelastic processes in the Eu~II + H~I collisions and the updated model atom of Eu~II. Non-LTE analyses of the Eu~II 4129 and 6645~\AA\ lines in spectra of the Sun and selected stars are performed in Section~3. Calculations of the non-LTE abundance corrections for the Eu~II lines are presented in Section~4. Section~5 formulates conclusions.

\section{Method of non-LTE calculations for Eu~II}\label{sect:method}

\subsection{Model atom of Eu~II}

Let us briefly remind the description of the Eu~II model atom, developed earlier \citep{eunlte2000}, and note the use of new atomic data.

{\bf Energy levels.} The atomic model is based on all 163 levels of Eu~II, presented in NIST. Closely spaced levels of the same parity are combined into superlevels. Fine splitting is taken into account for the terms \eu{5d}{9}{D}{\circ}{}, \eu{5d}{7}{D}{\circ}{}, and \eu{z}{9}{P}{}{}. The highest Eu~II level with \Eexc\ = 8.272~eV was obtained by combining \eu{8s}{9}{S}{\circ}{} and \eu{8s}{7}{S}{\circ}{}. The final system of 63 levels of Eu~II is closed by the ground state of Eu~III with the statistical weight $g = 8$. The excited states of Eu~III lie above 3.5~eV, and we neglect their contribution to the partition function, since in the range of stellar parameters of our interest the number density of Eu~III in the atmosphere is at least an order of magnitude less that for Eu~II.

{\bf Radiative rates.} For every of 109 bound-bound (b-b) transitions in the atomic model, we use oscillator strengths ($gf$-values) obtained by Kurucz (2012) in calculations of the Eu~II atomic structure. The radiative b-b rates are calculated with the Voigt absorption profile for the six strongest transitions from the ground and low-excited states and  with the Doppler profile for the remaining transitions. For Eu~II, there are no calculations of photoionization cross sections, and we apply the hydrogenic approximation with an effective principal quantum number instead of the principal
quantum number.

{\bf Collisional rates.} The electron-impact excitation rates are calculated using the van Regemorter formula (1962) if the transition is allowed, and with the effective collision strength $\Upsilon$ = 1 if the transition is forbidden. The electron-impact ionization rates are calculated using the Seaton (1962) formula and the adopted threshold photoionization cross sections.

To take into account inelastic processes in collisions with H~I atoms, we use data obtained in two different approaches. For 259 b-b transitions and 24 bound-free (b-f) transitions, the rate coefficients were calculated by one of the authors of this paper (S.A.Ya.). The characteristics of inelastic processes were calculated using the asymptotic approach within the Born-Oppenheimer approximation proposed by Belyaev (2013). This approach allows one to model the nonadiabatic regions associated with the ionic-covalent interaction of molecular terms, which in many cases is the dominant mechanism of inelastic processes occurring in collisions with hydrogen atoms \citep{Guitou2011}. This approach describes the ionic-covalent interaction by the off-diagonal matrix element of the electronic Hamiltonian that is calculated using the semi-empirical formula for charge exchange from \cite{Olson1971}. Such modeling of the electronic structure of a quasimolecule allows one to include into consideration only states of the same molecular symmetry as possessed by the ionic term. The scattering channel Eu$^{2+}(4f^7\,^{8}S)$ + H$^{-}(^{1}S)$ generates molecular states of $^8\Sigma^-$ symmetry, therefore only the Eu$^{+}(^{7,9}L)$ + H$(^{2}S)$ states which possess $^8\Sigma^-$ symmetry are included into consideration. The requirements of symmetry thus restrict the number of states, and a large number of the Eu~II states, for which the electron configuration and quantum numbers of the total angular momentum L and total spin S of electrons are not determined, is not included into consideration of inelastic hydrogen collision. In total, our calculations take into account 18 scattering channels, including one ionic term, for which the asymptotic energies are taken from the NIST database \citep{nist} and averaged over the quantum number J. Non-LTE calculations performed with this data set are called variant Y25. For the Eu~II levels that are included into model atom with the fine splitting, the rate coefficients are divided proportionally to the number of states in the upper level.

The second data set is taken from \cite{Storm2024}. These are the rate coefficients for 934 b-b transitions and 23 b-f transitions. The collisions of europium ions with hydrogen atoms were calculated using the LCAO (linear combination of atomic orbitals) approach \citep{Barklem2016}, which, like our approach, takes into account the ionic-covalent interaction in electronic structure modeling, and uses the multichannel Landau-Zener approach for the nuclear dynamics calculations. The difference of the LCAO approach from our study lies in the use of atomic orbitals for electronic structure calculations, not the semi-empirical expression for the matrix element, and in the fact that each scattering channel is coupled with a particular ionic term, that has the same molecular symmetry and the same core, leading to a larger number of ionic states of Eu$^{2+}$ + H$^{-}$ included into consideration. \cite{Storm2024} also added the rate coefficients calculated using the free electron model \citep{Amarsi2018} to the LCAO collision rates. This model allows one to calculate the rates for transitions between highly excited Rydberg states, and \cite{Amarsi2018} suggest that it takes into account contributions from mechanisms other than ionic-covalent interaction. It should be noted that a significantly larger number of Eu$^{+}$ + H states were taken into account in \cite{Storm2024} comparing to the calculations of S.A.Ya., but it was not explained how the terms with undetermined quantum numbers L and S were included. The non-LTE calculations performed with the data set from \cite{Storm2024} are called variant SBY24.

\begin{figure}  %[htbp]
	\hspace{-3mm}
	\centering
	\includegraphics[width=0.50\columnwidth,clip]{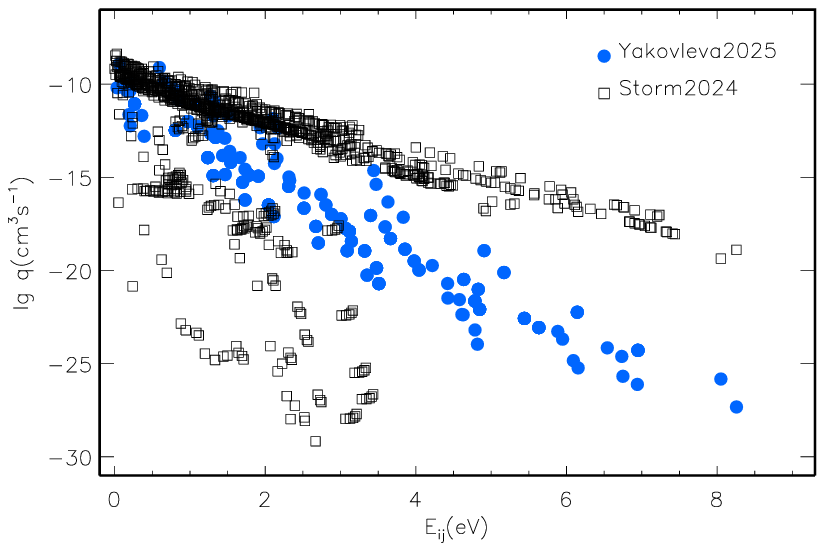}
	%\hspace{-8mm} 
	\vspace{-5mm}
	\caption{Excitation rate coefficients log~q~(cm$^3$s$^{-1}$) for the Eu~II + H~I collisions from calculations by S.A. Yakovleva (circles) and \citet[][squares]{Storm2024} as a function of the transition energy E$_{\rm ij}$. The data correspond to $T = 5000$~K. }
	\label{fig:coll}
\end{figure}

The excitation rate coefficients calculated by S.A.Ya. and \cite{Storm2024} are compared in Fig.~1. For some transitions the data of \cite{Storm2024} lie below the results of S.A.Ya., that is connected with the multichannel effect in the LCAO calculations. While the data from \cite{Storm2024} that lie above the data of S.A.Ya. correspond to the transitions for which the free electron model impact is large.

\subsection{Codes and Model Atmospheres}

To solve the system of the statistical equilibrium and radiative transfer equations in a given atmospheric model, we used the code {\sc detail} \citep{giddings81,butler84} with the updated opacity package \citep{mash_fe}. This study is based on plane-parallel, LTE model atmospheres, calculated in the MARCS project\footnote{http://marcs.astro.uu.se} (Gustafsson et al. 2008). The solution of an incomplete non-LTE problem is justified when studying the formation of Eu~II lines. 
Previously it was shown that for the Sun and solar-type stars accounting for the departures from LTE influences only a little on the atmospheric structure and radiation field \citep{Short2005}.

The europium abundances were determined from each individual line by automatically fitting the theoretical spectrum to the observed one. The synthetic spectra were calculated using the code \textsc{synthV\_NLTE} (Tsymbal et al., 2019), integrated into the BinMag visualization program \citep{Kochukhov_binmag}. Using b-factors (b = $n_{\rm NLTE} /n_{\rm LTE}$) from {\sc detail}, the code \textsc{synthV\_NLTE} calculates lines of Eu~II taking into account the non-LTE effects, while lines of the remaining elements in the LTE approximation. Here, $n_{\rm NLTE}$ and $n_{\rm LTE}$ are level populations, obtained by solving the statistical equilibrium equations (NLTE) and using the Boltzmann-Saha formulas (LTE).
 The line atomic parameters for calculating the synthetic spectra were taken from the recent version of the VALD3 (Vienna Atomic Line Database) database \citep{vald_hfs}. 
 
 \section{Analysis of Eu~II lines in spectra of the Sun and selected stars}\label{sect:sample}

 In this section, we test in which line formation scenario it is possible to achieve abundance agreement across different Eu~II lines for each of the sample stars. Four scenarios were considered: LTE, non-LTE (variant Y25), non-LTE (variant SBY24), and non-LTE (variant \kH\ = 0.1). The last option, in which collisions with hydrogen atoms are treated using the formulas of Steenbock and Holweger (1984) with a scaling factor of \kH\ = 0.1, will allow us to estimate the abundance errors due to applying the Drawin approximation. Abundances are determined from
two lines, Eu~II 4129~\AA\ (transition \eu{6s}{9}{S}{\circ}{4} -- \eu{z}{9}{P}{}{4}) and Eu~II 6645~\AA\ (transition \eu{5d}{9}{D}{\circ}{6} -- \eu{z}{9}{P}{}{5}), which are subject to non-LTE effects of the opposite sign. In non-LTE calculations, the Eu~II 4129~\AA\ line is always weaker compared to LTE, while Eu~II 6645~\AA\ can be enhanced under certain atmospheric parameters. Oscillator strengths, log~$gf$ = 0.22 (4129~\AA) and 0.12 (6645~\AA), and parameters of the hyperfine splitting (HFS) components of the isotopes $^{151}$Eu and $^{153}$Eu  were obtained by \cite{Lawler_Eu} in laboratory measurements.

\subsection{Stellar sample, observations and atmospheric parameters}

The Sun is a reference star with an effective temperature of \Teff\ = 5780~K, surface gravity of \lgg\ = 4.44 and microturbulence velocity of $\xi_{t}$ = 0.9~\kms. We use the spectrum of the Sun as a star, obtained with a spectral resolving power of R = $\lambda/\Delta\lambda \simeq$ 300\,000 in blue and R $\simeq$ 520\,000 in the red spectral range (Kurucz et al., 1984).

\begin{table}
\centering
\renewcommand{\arraystretch}{1.0}
\renewcommand{\tabcolsep}{7pt}
\caption{Stellar sample, atmospheric parameters, LTE and non-LTE abundances derived from lines of Eu~II. }
\vspace{3mm}
\label{tab:stars}
\begin{tabular}{lcccccc}
	\hline \noalign{\smallskip}
	Star, HD & Sun &   8724 &  83212 &  122956 &  165195 &  204543  \\
	\noalign{\smallskip} \hline \noalign{\smallskip}
	\Teff\ [K]  & 5777   & 4560   & 4550   & 4720   & 4420   & 4670   \\
	\lgg       & 4.44   & 1.29   & 1.00   & 1.53   & 0.97   & 1.41   \\
${\rm [Fe/H]}$ & 0      & --1.71 & --1.44 & --1.67 & --2.15 & --1.59 \\
	$\xi_t$ [\kms] & 0.9 &    1.5 &    1.9 &    1.8 &    2.0 &   1.9  \\
 &	\multicolumn{6}{l}{~~\ion{Eu}{2} 4129~\AA: $\eps{}$ (dex)} \\
	LTE        & 0.52 & --0.96 & --0.65 & --0.77 & --1.36 & --1.01  \\
	SBY24      & 0.57 & --0.81 & --0.58 & --0.68 & --1.24 & --0.91 \\
	Y25        & 0.60 & --0.76 & --0.55 & --0.62 & --1.17 & --0.84 \\
	\kH\ = 0.1 & 0.58 & --0.80 & --0.58 & --0.67 & --1.22 & --0.90 \\
 &	\multicolumn{6}{l}{~~\ion{Eu}{2} 6645~\AA: $\eps{}$ (dex)} \\
	LTE        & 0.61 & --0.81 & --0.47 & --0.70 & --1.20 & --0.86 \\
	SBY24      & 0.62 & --0.83 & --0.47 & --0.74 & --1.23 & --0.91 \\
	Y25        & 0.62 & --0.84 & --0.47 & --0.75 & --1.23 & --0.91 \\
	\kH\ = 0.1 & 0.60 & --0.86 & --0.48 & --0.77 & --1.26 & --0.93 \\
 &	\multicolumn{6}{l}{~~Mean non-LTE (SBY24)} \\
${\rm [Eu/Fe]}$ & 0.08 &  0.38 & 0.41 & 0.45 & 0.40 & 0.17  \\
\multicolumn{1}{c}{$\sigma$} & 0.04 &  0.01 & 0.08 & 0.04 & 0.01 & 0.00  \\
	\noalign{\smallskip}\hline
\end{tabular}
\end{table}

\begin{figure}  %[htbp]
	%\hspace{-6mm}
	\centering
	\includegraphics[width=0.45\columnwidth,clip]{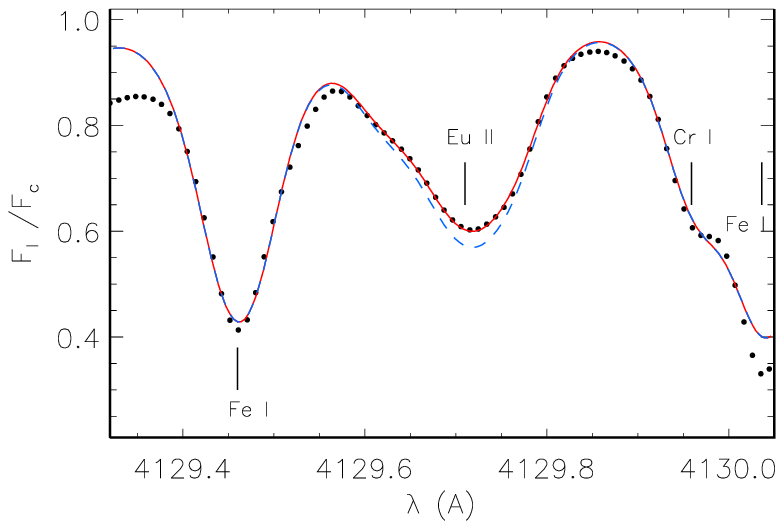}
	\includegraphics[width=0.45\columnwidth,clip]{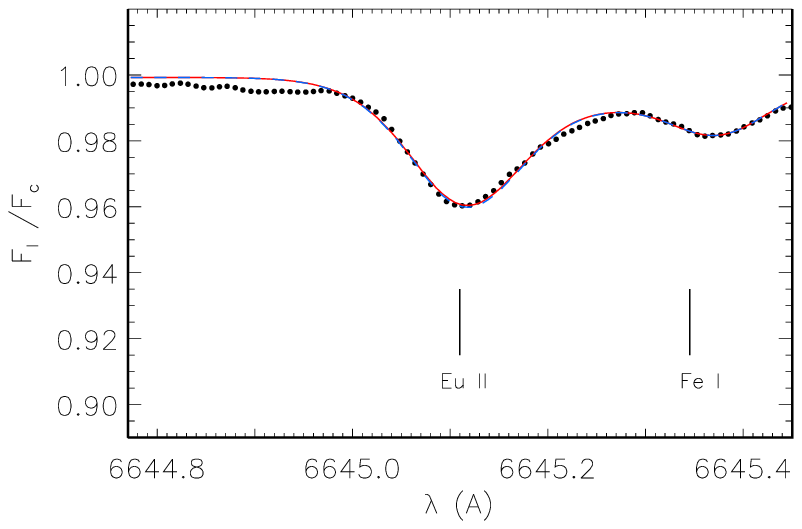}
	\vspace{4mm}
	\includegraphics[width=0.45\columnwidth,clip]{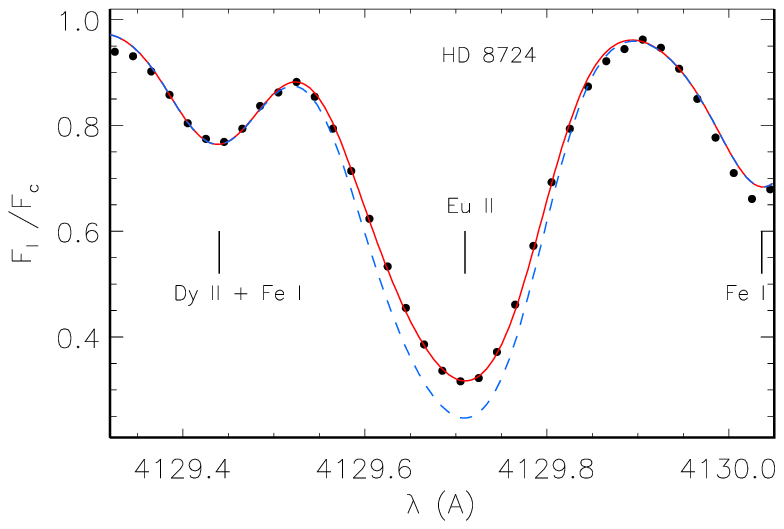}
    \includegraphics[width=0.45\columnwidth,clip]{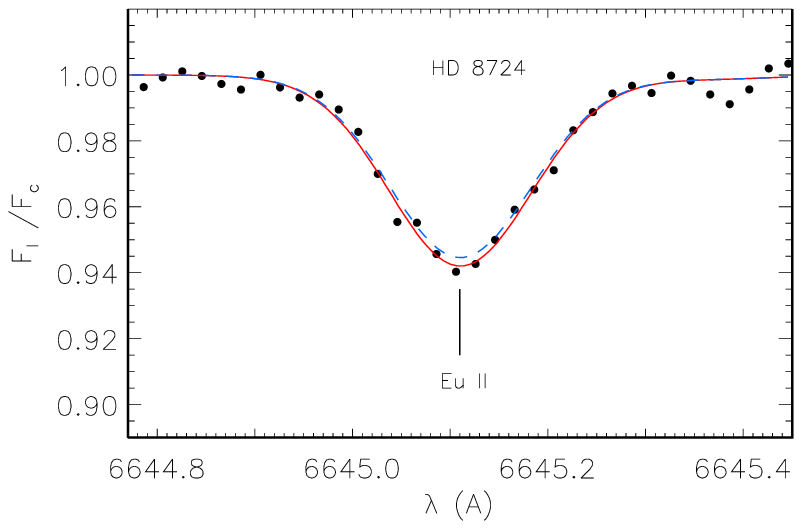}
	%\hspace{-8mm}
	\vspace{-8mm}
	\caption{Observed line profiles (circles) of \ion{Eu}{2} 4129 and 6645~\AA\ in spectra of the Sun (top row) and HD~8724 (bottom row) in comparison with synthetic spectra, calculated taking into account non-LTE effects (SBY24, solid curve) for lines of Eu~II. The obtained non-LTE abundances are given in Table~\ref{tab:stars}. The dashed curve in each panel corresponds to the LTE calculations with the Eu abundance obtained at non-LTE.
}
	\label{fig:eu2}
\end{figure}

Of all the stars for which we had previously determined the atmospheric parameters, we were able to select only six stars with [Fe/H] $< -1$, where both Eu~II lines are reliably measured. All of them are from the study of \citet{Mashonkina_pd}, who describe in detail the determinations of atmospheric parameters and the
characteristics of the observed spectra. Here we only note that the effective temperatures are determined by modern photometric methods, \lgg\ are based on the Gaia parallaxes (Gaia collaboration, 2021), stellar spectra with R $> 40\,000$ were taken from the archive of the European Southern Observatory, namely, Very Large Telescope (VLT2). The sample stars and their atmospheric parameters are listed in Table~1.

\subsection{Determination of europium abundances}

The best non-LTE fits of \ion{Eu}{2} 4129~\AA\ and 6645~\AA\ in the spectra of the Sun and HD~8724 are shown in Fig.~2. Abundances $\eps{}$, obtained for all stars from individual lines in different line formation scenarios, are given in Table~1. We use the abundance scale where $\eps{H}$ = 12. The [Eu/Fe] values were calculated using the mean non-LTE (SBY24) abundances from two lines and the solar abundances $\eps{\odot,Eu}$ = 0.51 and $\eps{\odot,Fe}$ = 7.45 \citep{lodders21}. The statistical abundance error $\sigma$ is calculated as the dispersion in the single-line measurements around the mean.

For the Sun, the LTE abundance from Eu~II 4129~\AA\ is lower than from the second line, by 0.09~dex. The difference in non-LTE abundances does not exceed 0.05~dex because of different effects on these lines. For Eu~II 4129~\AA, the difference between non-LTE and LTE abundances (non-LTE correction $\Delta_{\rm NLTE} = \eps{NLTE} - \eps{LTE}$) is  $\Delta_{\rm NLTE}$ = 0.05~dex in the SBY24 variant and 0.08~dex in the Y25 variant, while $\Delta_{\rm NLTE}$ does not exceed 0.01~dex for \ion{Eu}{2} 6645~\AA. Due to weak non-LTE effects, the solar Eu~II lines are little suitable for testing the atomic model.

In stars with [Fe/H] $< -1.5$, non-LTE in all variants leads to weakened Eu~II 4129~\AA\ line with $\Delta_{\rm NLTE}$ of 0.09 to 0.19~dex, but strengthened \ion{Eu}{2} 6645~\AA\ line with negative non-LTE correction (Table~1). Different sign of $\Delta_{\rm NLTE}$ for Eu~II 4129 and 6645~\AA\ in one model atmosphere can be understood from  the behavior of the departure coefficients (b-factors) of the lower and upper levels in the transitions between which these lines are formed. Figure~3 shows the b-factors of the \eu{6s}{9}{S}{\circ}{4} and \eu{z}{9}{P}{}{4} levels, associated with Eu~II 4129~\AA, and \eu{5d}{9}{D}{\circ}{6} and \eu{z}{9}{P}{}{5} levels, associated with 6645~\AA, in the model atmosphere of the star HD~204543. Since Eu~II is the dominant ionization stage, then the b-factor of the ground state (\eu{6s}{9}{S}{\circ}{4}) is close to 1 at all depths in the atmosphere. In the line formation region, where the optical depth at a wavelength of 5000~\AA\ log~$\tau_{5000} < 0.6$, all excited levels are overpopulated relative to thermodynamic equilibrium populations (b $> 1$), due to radiative pumping of the transitions from the ground state. Therefore, for Eu~II 4129~\AA\ in the region of its formation (log~$\tau_{5000} \simeq -1.4$), the line source function $S_\nu \simeq B_\nu(T) \cdot$ b(\eu{z}{9}{P}{}{4})/b(\eu{6s}{9}{S}{\circ}{4}) exceeds the Planck function, and the line is weaker than in the LTE calculations. Such an inequality $S_\nu > B_\nu(T)$ is also true for Eu~II 6645~\AA\ in the region of its formation (log~$\tau_{5000} \simeq -0.1$),
and this effect tends to weaken the line compared to LTE. But the lower level (\eu{5d}{9}{D}{\circ}{6}) of the transition is overpopulated, which makes the line stronger compared to LTE. In the competition between two effects, the second one wins.

\begin{figure}  %[htbp]
	\hspace{-3mm}
	\centering
	\includegraphics[width=0.50\columnwidth,clip]{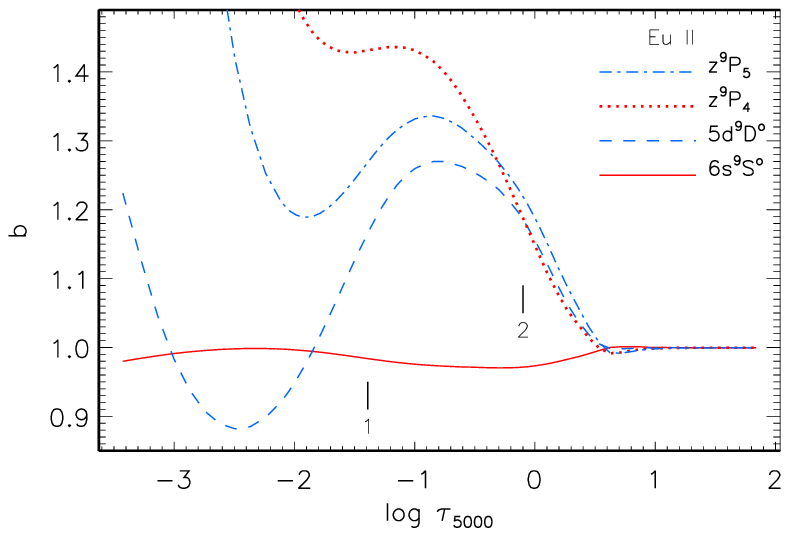}
		%\hspace{-8mm}
	\vspace{-5mm}
	\caption{b-factors of the \ion{Eu}{2} selected levels as a function of log~$\tau_{5000}$ in the atmospheric model with \Teff\ = 4670~K, \lgg\ = 1.41, [Fe/H] = $-1.58$. The tick marks indicate the locations of line center optical depth unity for the \ion{Eu}{2} 4129~\AA\ (1) and 6645~\AA\ (2) lines.}
	\label{fig:bf}
\end{figure}

\begin{figure}  %[htbp]
	%\hspace{-6mm}
	\centering
	\includegraphics[width=0.45\columnwidth,clip]{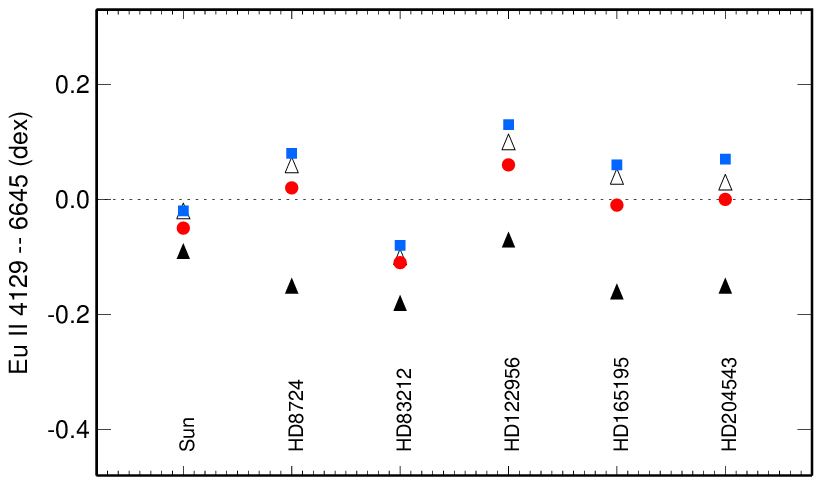}
%	\includegraphics[width=0.45\columnwidth,clip]{astars_N.pdf}
	%\hspace{-8mm}
	\vspace{-5mm}
	\caption{Abundance differences between Eu~II 4129~\AA\ and 6645~\AA\ in different line formation scenarios: LTE (filled triangles), non-LTE (SBY24, circles), non-LTE (Y25, squares), and non-LTE (\kH\ = 0.1, empty triangles)    for our sample stars. }
	\label{fig:stars}
\end{figure}

Figure~4 shows the abundance differences between Eu~II 4129~\AA\ and 6645~\AA\ in different line formation scenarios for all our sample stars. The LTE abundance from the first line is systematically lower than that from the second line, by 0.07 to 0.18~dex. Any of the non-LTE calculations improves the situation, except perhaps for the star HD~122956, for which the difference of 0.13~dex is greater in the Y25 case than in the LTE case. Obviously, accounting for the deviations from LTE leads to decreasing the abundance determination errors. For example, for HD~165195 $\sigma$ = 0.01~dex in the SBY24 variant, while $\sigma$ = 0.11~dex in the LTE case.

When determining the Eu abundances of stellar samples covering a wide metallicity range (up to [Fe/H] $\sim -3$),  the Eu~II 4129~\AA\ and 4205~\AA\ resonance lines are used, as a rule. Our calculations for Eu~II 4129~\AA\ in the $-2.15 \le$ [Fe/H] $\le 0$ range show that the abundance difference between the SBY24 and \kH\ = 0.1 variants is small, of 0.00 to 0.02~dex for different stars. This means that our earlier non-LTE abundance determinations for  europium \citep[][and others]{Zhao2016,mash_dsph,Mashonkina_pd} do not require significant revision.

\begin{table}
	\centering
	\renewcommand{\arraystretch}{1.0}
	\renewcommand{\tabcolsep}{7pt}
	\caption{Eu~II lines, for which a grid of non-LTE abundance corrections was calculated.}
	\vspace{3mm}
	\label{tab:lines}
	\begin{tabular}{ccr}
		\hline \noalign{\smallskip}
		$\lambda$ (\AA) & \Eexc\ (eV)  & log $gf$  \\
		\noalign{\smallskip} \hline \noalign{\smallskip}
3724.9 &  0.00  & --0.09  \\
3819.6 &  0.00  &   0.51  \\
3907.1 &  0.21 &   0.17  \\
3930.5 &  0.21 &   0.27  \\
3971.9 &  0.21 &   0.27  \\
4129.7 &  0.00  &   0.22  \\
4205.0 &  0.00  &   0.21  \\
4435.5 &  0.21 & --0.11  \\
4522.5 &  0.21 & --0.67  \\
6437.6 &  1.32 & --0.32  \\
6645.0 &  1.38 &   0.12  \\
 	\noalign{\smallskip}\hline
\end{tabular}
\end{table}

For the Sun, the mean from the two lines $\eps{Eu}$ = 0.59$\pm$0.04 (SBY24) or 0.61$\pm$0.01 (Y25) is above the  meteoritic abundance $\eps{met}$ = 0.51$\pm$0.02 \citep{lodders21}. Using only Eu~II 6645~\AA, Storm et al. (2024) recommend $\eps{\odot}$ = 0.57 based on the 3D NLTE calculations. Thus, taking into account the 3D effects did not allow Storm et al. to reconcile the solar abundance with the meteoritic one. Our results are consistent with the 1D NLTE abundance $\eps{\odot}$ = 0.60 obtained by Storm et al. (2024).

\begin{table}
	\centering
	\renewcommand{\arraystretch}{1.0}
	\renewcommand{\tabcolsep}{7pt}
	\caption{Parameters of the model atmospheres with which the non-LTE abundance corrections for Eu~II lines were calculated.}
	\vspace{3mm}
	\label{tab:grid}
	\begin{tabular}{cccc}
		\hline \noalign{\smallskip}
 \ [Eu/Fe] & \Teff\ (K)  & \lgg & [Fe/H] \\
	\noalign{\smallskip} \hline \noalign{\smallskip}
 0.0 to 1.5 & 4000 to 4750 & 0.5 to 2.5 & --2.0 to --4.0  \\
 0.0 to 1.5 &         5000 & 0.5 to 5.0 & --2.0 to --4.0  \\
 0.0 to 1.5 &  5250, 5500  & 2.0 to 5.0 & --2.0 to --4.0  \\
 0.0 to 1.5 & 5750 to 6500 & 3.0 to 5.0 & --2.0 to --4.0  \\
 0.0,  0.5  & 5000 to 6500 & 3.0 to 5.0 & ~~0.0 to --1.5  \\
 --0.5      & 4000 to 4750 & 0.5 to 2.5 & --2.0 to --4.0  \\
 	\noalign{\smallskip}\hline
 \end{tabular}
\end{table}

\section{Non-LTE abundance corrections for Eu~II lines}\label{sect:NLTEcorr}

A new atomic model (variant SBY24) was used to calculate non-LTE abundance corrections for 11 Eu~II lines, listed in Table~2. The oscillator strengths and the HFS parameters were taken from the measurements of Lawler et al. (2001). The calculations were carried out with the MARCS model atmospheres (Gustafsson et al., 2008) for a set of atmospheric parameters characteristic of stars that are used in studies of the galactic chemical evolution: 4000~K $\le$ \Teff\ $\le$ 6500~K, 0.5 $\le$ \lgg\ $\le$ 5.0, $-4.0 \le$ [Fe/H] $\le$ 0.0. Microturbulent velocity is taken equal to $\xi_t$ = 2~\kms\ for models with \lgg\ $\le$ 2.5 and 1~\kms\ for the rest. Steps by \Teff, \lgg, and [Fe/H] are 500~K, 0.5, and 0.5, respectively. The MARCS website does not contain models with [Fe/H] = $-3.5$, so for \Teff\ $\le$ 5000~K and \lgg\ $\le$ 2.5 they were obtained by interpolation between models with [Fe/H] = $-3.0$ and $-4.0$ using the algorithm presented at MARCS website. For other \Teff /\lgg\ parameters and [Fe/H] = $-3.5$, equivalent width (EW) of the strongest Eu~II lines does not exceed 1~m\AA\ even in the [Eu/Fe] = 1.5 case. Non-LTE calculations were performed with different europium abundances: [Eu/Fe] = $-0.5$, 0.0, 0.5, 1.0, and 1.5. Thus, for each individual line, the 4-dimensional grid of non-LTE corrections was built. However, non-LTE calculations were not performed for every grid node, see Table~3. For example, in the [Fe/H] $< -2$ range, there is a large spread in stellar Eu abundances, with [Eu/Fe] from $-0.5$ to almost 2, but negative [Eu/Fe] values cannot be associated with dwarf stars, since the Eu~II 4129~\AA\ line is too weak in their spectra to be measured. Therefore, calculations with [Eu/Fe] = $-0.5$ were only produced for giants (0.5 $\le$ \lgg\ $\le$ 2.5) with \Teff\ from 4000 to 4750~K in the [Fe/H] $\le -2.0$ range. In the same range, stars with a very high europium abundance ([Eu/Fe] $> 1$, [Eu/Ba] $> 0$) are observed. According to Christlieb et al. (2004), they belong to the r-II group and play a key role in obtaining the empirical r-process element abundance pattern (Sneden et al., 2008). Abundances of Eu and other heavy elements in the r-II stars should be determined with the highest possible accuracy. Therefore, for the [Fe/H] $\le -2.0$ range, calculations were carried out with [Eu/Fe] = 1.0 and 1.5.

\begin{figure}  %[htbp]
	%\hspace{-6mm}
% \vspace{-1cm}
	\centering
	\includegraphics[width=0.45\columnwidth,clip]{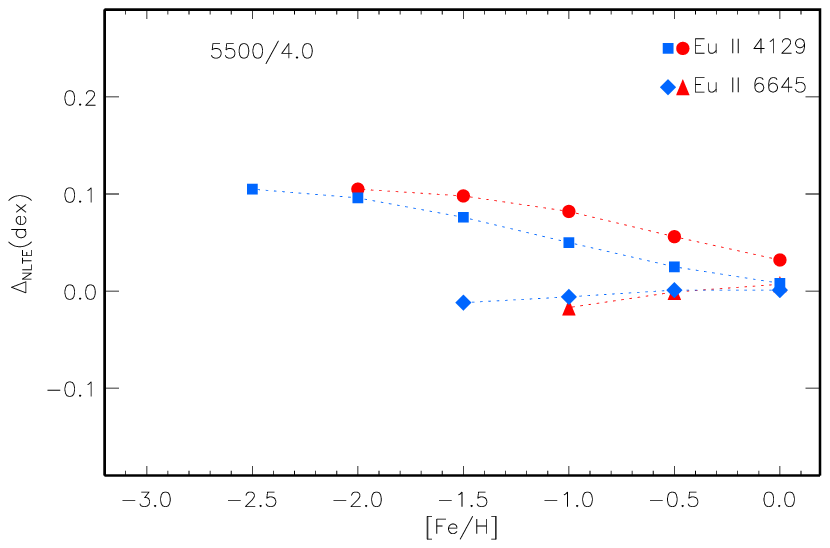}
	\includegraphics[width=0.45\columnwidth,clip]{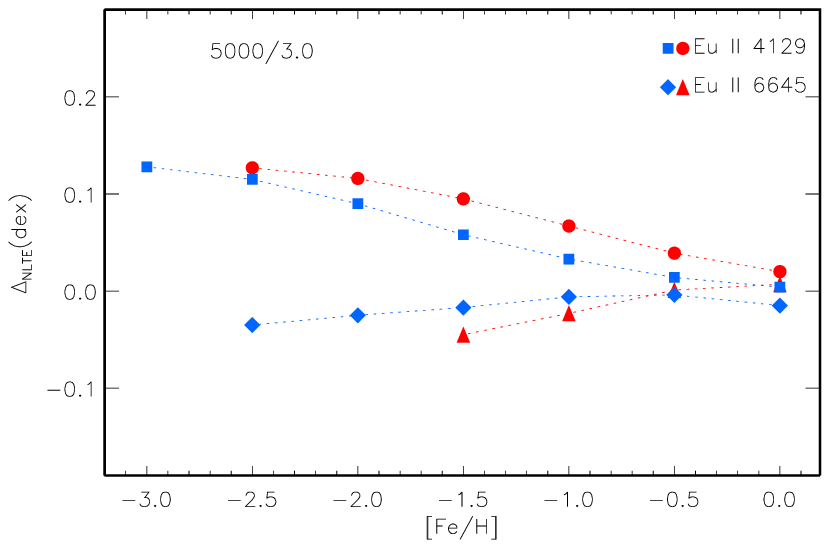}
	%\hspace{-8mm}
	\vspace{-5mm}
	\caption{Non-LTE abundance corrections for Eu~II 4129~\AA\ and 6645~\AA\ as functions of metallicity for two pairs of \Teff /\lgg\ = 5500~K/4.0 (left panel) and 5000~K/3.0 (right panel). Circles and triangles (red symbols) correspond to calculations with [Eu/Fe] = 0; squares and diamonds (blue symbols) correspond to
calculations with [Eu/Fe] = 0.5.}
	\label{fig:dnlte_dwarf}
\end{figure}

 \begin{figure}  %[htbp]
 %\hspace{-6mm}
 %\vspace{-1cm}
 	\centering
 	\includegraphics[width=0.45\columnwidth,clip]{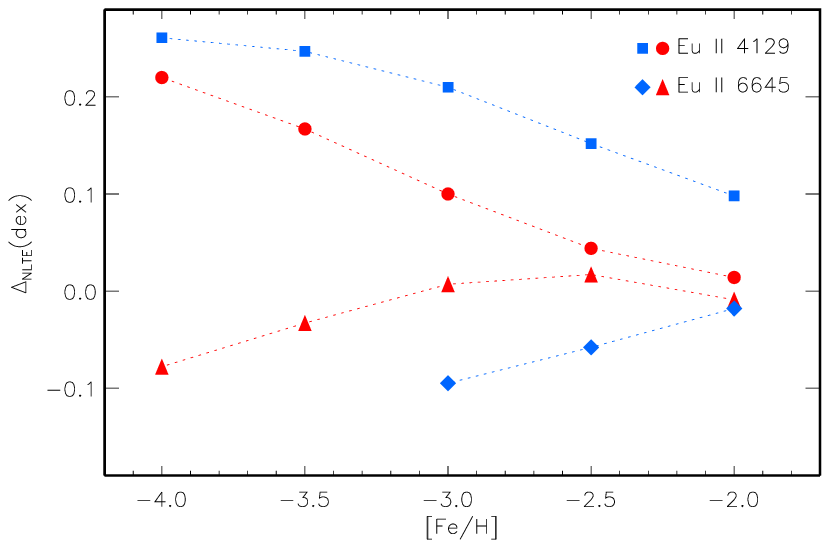}
 %\hspace{-8mm}
 \vspace{-5mm}
 	\caption{The same as in Fig.~\ref{fig:dnlte_dwarf} for model atmospheres of giants with \Teff\ = 4500~K, \lgg\ = 1.5. Squares and diamonds (blue symbols) correspond to calculations with [Eu/Fe] = 0.5; circles and triangles (red symbols) for calculations with [Eu/Fe] = 1.5. }
% 	Расчеты сделаны с вариантом SBY24 модели атома. }
 	\label{fig:dnlte_giant}
 \end{figure}

Non-LTE corrections and equivalent widths are posted on the website INASAN\_NLTE5\footnote{\tt http://spectrum.inasan.ru/nLTE2/}, where $\Delta_{\rm NLTE}$ for a given line(s) and given \Teff, \lgg, [Fe/H] and [Eu/Fe] can be obtained {\it online} by interpolation. Figures~\ref{fig:dnlte_dwarf} and \ref{fig:dnlte_giant}
demonstrate the dependence of non-LTE corrections for the Eu~II 4129~\AA\ and 6645~\AA\ lines on metallicity and
europium abundance in the models representing the atmospheres of main sequence (\lgg\ = 4), turn-off (\lgg\ = 3) and giant (\lgg\ = 1.5) stars. For any pair of \Teff /\lgg, the non-LTE corrections are positive for Eu~II 4129~\AA, but negative or close to 0 for Eu~II 6645~\AA, and for both lines they increase in absolute value with decreasing [Fe/H]. The increase in deviations from LTE with decreasing metal abundance is explained by a decrease in the electron number density, and hence a weakening of the role of collisional processes in establishing statistical equilibrium of the europium ion, and a decrease in opacity in the ultraviolet range, and hence an increase in the role of radiative processes. For the same set of \Teff /\lgg /[Fe/H], the non-LTE effects are weaker in the model with higher [Eu/Fe]. This can be explained by the fact that at higher [Eu/Fe] the lines are stronger and the line wings formed in the deep atmospheric layers make a greater contribution to the line absorption. The use of {\it correct} element abundance in the non-LTE calculations is especially important for the [Fe/H] $< -2$ range, where a large spread of [Eu/Fe] values is observed.

{\it Comparison with the literature data.} Our correction $\Delta_{\rm NLTE}$ = +0.01~dex for the Eu~II 6645~\AA\ solar line (variants SBY24 and Y25) coincides with the result of Storm et al. (2024) in the case of using the same atmospheric model (MARCS) as in our work.

Guo et al. (2025) used the non-LTE method from Storm et al. (2024) and calculated the non-LTE corrections for Eu~II 4129~\AA\ and 6645~\AA\ in the MARCS models with \Teff /\lgg\ = 4500/1.5, 5500/3.5, and 6000/4.0 for different [Fe/H] and one value of [Eu/Fe] = 0 (see their Fig.~4). In our non-LTE calculations with the same atmospheric parameters, EW(Eu~II 4129\AA) becomes less than 1~m\AA\ for [Fe/H] $< -3.5$ at 4500/1.5, [Fe/H] $< -2$ at 5500/3.5, and [Fe/H] $< -1.5$ at 6000/4.0, so we do not compare corrections for these ranges. For all models with [Fe/H] = $-2$, our $\Delta_{\rm NLTE}$(4129~\AA) are consistent within 0.02~dex with the results of Guo et al. (2025). In the models with [Fe/H] = 0 and $-1$, we obtained slightly smaller non-LTE effects for both lines than Guo and et al. (2025), so for the 5500/3.5 models our $\Delta_{\rm NLTE}$(4129~\AA) is 0.04~dex and 0.03~dex smaller and $\Delta_{\rm NLTE}$(6645~\AA) is less negative by 0.02~dex and 0.01~dex, respectively.

\section{Conclusions}\label{sect:conclusions}

 The Eu~II model atom, built by Mashonkina (2000) and \cite{eunlte2000}, was updated by implementing the rate coefficients for the excitation and ion pair formation in the Eu~II + H~I collisions, calculated by Storm et al. (2024) (variant SBY24) and in this paper (variant Y25). To test the atomic model, calculations were
performed with different scenarios of line formation for Eu~II in classical model atmospheres of the Sun and five stars with reliably determined atmospheric parameters, and for each scenario the abundances were determined from the Eu~II 4129~\AA\ and 6645~\AA\ lines. The LTE abundance from the first line is systematically lower than that from the second one, by 0.09~dex for the Sun and by 0.07 to 0.18~dex for metal-deficient stars. Accounting for the non-LTE effects allows the abundances from the two lines to be reconciled within the determination error, since Eu~II 4129~\AA\ is weakened and the non-LTE abundance correction is positive, while Eu~II 6645~\AA, on the contrary, is strengthened and the non-LTE correction is negative. In different stars, the abundance difference between the two lines varies from 0.00 to 0.11~dex in absolute value in calculations with the SBY24 atomic model and from 0.02 to 0.13~dex with the Y25 atomic model. On average, the difference is smaller for the SBY24 than Y25 variant.

Using the synthetic spectrum method, we derived the solar non-LTE abundance $\eps{Eu}$ = 0.59$\pm$0.04 from the two lines, Eu~II 4129~\AA\ and 6645~\AA. It is 0.08~dex higher than the meteoritic abundance recommended by Lodders (2021). The discrepancy between the solar and meteoritic abundances was also found in the 3D NLTE calculations by Storm et al. (2024). Further research is required to understand the source of discrepancies. Accounting for the non-LTE effects for Eu~II reduces stellar abundance errors, to $\sigma \le$ 0.04~dex for 5 out of 6 stars in our sample.

A grid of non-LTE abundance corrections for 11 lines of Eu~II was calculated for a wide range of stellar parameters: 4000~K $\le$ \Teff\ $\le$ 6500~K, 0.5 $\le$ \lgg\ $\le$ 5.0, $-4 \le$ [Fe/H] $\le 0$,  $-0.5 \le$ [Eu/Fe] $\le$ 1.5. The non-LTE corrections increase in absolute value with decreasing \lgg\ and [Fe/H] and for the fixed set of \Teff /\lgg /[Fe/H] depend on the Eu abundance. The data are publicly available on the INASAN\_NLTE website and can be used in studies of the chemical evolution of galaxies to refine stellar europium abundances.

\acknowledgments
The authors acknowledge P.~S. Barklem for consultations and posting his results of Eu~II + H~I collision calculations in the github repository; Yu.~V. Pakhomov for maintaining the INASAN\_NLTE website. S.A.Ya. expresses gratitude to Professor A.~K. Belyaev. 
The work was carried out under the state assignment of the Institute of astronomy of RAS and with the support of an internal grant of the A.~I. Herzen State Pedagogical University of Russia (project 46-VG).
This study made use of the MARCS, NIST, VALD and ADS\footnote{http://adsabs.harvard.edu/abstract\_service.html} databases.

\end{document}